# Language Support for Declarative Future Commitments

William Harrison
Department of Computer Science
Trinity College
Dublin 2, Ireland*
(+353) 1-896 8556

Bill.Harrison@cs.tcd.ie

## ABSTRACT
Sequential programming and work-flow programming are two useful, but radically different, ways of describing computational processing. Of the two, it is sequential programming that we teach all programmers and support by programming languages, whether in procedural, object-oriented, or functional paradigms. We teach workflow as a secondary style of problem decomposition for use in special situations, like distributed or networked processing. Both styles offer complementary advantages, but the fact that they employ radically different models for ownership of continuations interferes with our ability to integrate them in a way that allows them to be taught and used in a single programming language. This paper describes a programming language construct, declarative future commitments, that permit better integration of the two.

## Categories and Subject Descriptors
D.2.11 [**Software Architectures**]: *Languages*

D.3.3 [**Language Constructs and Features**]: *Concurrent programming structures*

## General Terms
Languages, Design

## Keywords
event-flow, broadcast, complex-event-processing, event-driven programming

## 1. INTRODUCTION
Most programmers are taught to decompose software into modular units of function that are combined to solve an overall problem. These units may ordinarily be called methods, functions or procedures, or given other names. In this paper we will use the term *task* because we want to focus on the intentional side rather than on its embodiment as a device for controlling a sequence of subtasks. As we use the term, a task is the actual function to be performed, without reference to how that is encoded. In referring to its encoding, we will use the more usual terms method or procedure at random to avoid bias, or other encoding artifacts, as appropriate. Methods and procedures are connected in an essentially sequential manner by a call/return mechanism.

While any task or collection of tasks is actively executing or paused in its execution, there remain another set of tasks which might or will be performed in the future, perhaps dependent on the outcomes of the currently executing tasks. Formal models of computation often refer to the abstraction of this concept of "the meaning of the rest of the program" as a *continuation*[18]. The discussion of continuations in Section 2 relates three issues of interest to us in this paper: 1) who defines the larger body of work in which the active tasks and those in the continuation operate, 2) how do the programming languages offer control of the continuation to this definer, and 3) what are the resource overheads of the offered manner of making a request for task to be performed.

A few general words about these issues are in order before examining the common alternatives. The issue of who defines the larger body of work is often ignored in teaching software engineering. We tend to focus on the part we know that we want done. In particular, it is common for the initial stakeholders in the design of a software system to imagine that they are the only stakeholders, and that they know all the requirements imposed on the system. In the small, this idea manifests itself in the idea that the designer of a task will "use" other services or components to accomplish its known end. But the realities of modern software are that the developer of a task may be aware of only part of the function that the overall system requires. In calling for a sub-task to achieve some result, additional behavior may be triggered for other functional or systemic purposes. For example, the developer of a library filing system may imagine that the "addBook" method carries out the task of adding the book to the catalog, and be quite unaware that the librarian later intends it to have the additional behavior of notifying library members that a new book is available. The additional behavior plays no role in the method implementation's performance of the original purposes for which it was called. But, the overall purpose of the task, of any task, is composite, reflecting the needs not only of a client, but those of other clients and stakeholders as well. Supporting the use of a multiplicity of methods to carry out the overall purpose of a task, as defined by many stakeholders, was the underlying intent of subject-oriented[7] and later aspect-oriented[14] programming.

* This work is supported by a grant from Science Foundation, Ireland

For this reason, the issue of compositional programming must be part of the discussion of language support. Aspect-oriented languages[3], for instance, allow the weaving of additional behavior to effect composite behavior. But once we step back from the idea that the designer of an algorithm has control of what is happening "in the large", we can imagine that some or all of the work performed consequent to a call might be able to proceed with some degree of independence or even in parallel with the caller's expected continuation[8]. This takes us into the domain of events, and of language supporting both call/return and event-like interactions.

This use of call/return vs. events has implications for run-time overheads in supporting the mechanisms as well, and in the next section we shall examine the several mechanisms in use for managing control of continuations in terms of the trade-offs made in the areas of ownership, compositional effects and overhead, leading to the identification of a language feature that helps bridge the differences.

## 2. CONTINUATIONS AND COMPOSITION

### 2.1 Stack Models of Control Flow
   – Client Control of the Continuation

The stack model for control is probably the most widely know model for managing continuations. An executing procedure has control of the flow of work, and has access to state information retained in storage. When a call is made, the state of the variables and execution is remembered as part of the continuation, and execution starts in the called procedure[16]. As the nesting of calls gets deeper and deeper, the resources required to store this continuation grow, and can become quite substantial. When, perhaps because of resource scheduling or distribution, a call is made for which there is a long latency, the entire continuation must be stored until the called procedure finishes. At that time, the information from the continuation is restored and the client's execution is resumed, putting the writer of the client in control of what happens next. The writer of a procedure is generally not concerned with the cost of the state being stored, which includes not only that for the procedure itself, but that of its (unknown) client, recursively up the stack. If the called procedure's execution is slow or delayed, the continuation's potentially large storage can be tied up for quite a long time. This resource consumption is one of the penalties paid for using this model to give control of the continuation to the client.

The stack model for execution has not encouraged compositional programming, and aspect-oriented technologies have generally addressed this difficulty through in-line weaving of material from the various aspects. Early technologies[7] exploited method calls as natural points for joining aspects, since the client was in any case relinquishing control to an unknown service. Modern technologies[13] allow the service to use more complex predicates to describe other points at which behavior can be added to previously written code even when those points had not been encapsulated into method calls by their developers.

### 2.2 Work-flow Models of Control Flow
   – Outsider Control of the Continuation

The workflow model presents an entirely different picture. In the workflow view, the tasks do not call upon one another but play a role in the orchestration defined by the workflow designer. Frequently, the tasks are connected by *messages*. Each task may proceed when its triggering message is received, and each task produces messages as part of its output. The workflow designer describes how the messages are routed among the tasks. Perhaps the simplest workflow language is the system of pipes provided by the UNIX system. Each task receives messages from its input pipe and produces new messages its output pipe(s)[1]. Workflow systems and their languages have not been formalized or systematized by computer scientists to the same extent as programming languages. Most of them are graphical with notation and detailed interpretation specified by manufacturers or by application domain groups. An excellent summary of the underlying principles as well as the chaotic product landscape can be found in [21] and a simplified explanation in [12].

Unlike the stack model, a task does not orchestrate the other tasks used to process the messages it sends. The task is not called from a client, so the continuation state is not made up by an indefinite stack of clients. All of the tasks are controlled from the workflow itself, and the continuation's state is held in the messages yet to be delivered, and heap objects to which they refer.

There is no difference in computing power between workflow and stack models, so a given set of tasks could be implemented with equivalent efficiency in either. But different choices about representation and expressability can lead to rather different characteristics. For example, messages might be restricted to reference only objects in reliable storage, like files, although access may be slower. Those objects may themselves be similarly restricted. This may result in less accidental retention of unneeded or redundant state in a continuation, but might cause the developer more effort when information needs to be retained for future tasks.

---

[1] Strictly speaking, the messages are single characters, since the UNIX model involves character streams.

The work-flow model provides a more natural base for compositional development than does the stack model. It exposes the tasks directly at the top level, highlighting their information dependencies, rather than hiding them within the flow of other tasks and it emphasizes their role "in the large" rather than the use to which some other task puts them.. The task provides a natural join-point for attachment of other behavior, avoiding the need for special languages for attaching aspects.

However, the converse of this property may be the most significant drawback to the workflow model. Programmers using available workflow systems code the low-level tasks using a stack-model language. Thus, the workflow itself can be relegated to use for a special class of problems, and treated not as a way of thinking about problems in general but as a technique applicable to this class of problems.

The alternative of entirely abandoning stack-model programming languages does not appear viable. Many attempts to encourage the use of what are called "visual programming languages," including dataflow languages, have been made. The have not been adopted in a widespread fashion. The reasons are probably a mixture of psychological reasons like those discussed in Section 2.1 with historical ones. Historically, programming languages have traditionally used a linear syntax that allows a program to be expressed as a character stream, even though devices that facilitate more readable forms of expression have been commonly available for over 20 years. This linear form also provides a standard form for interchanging program material; but alternative technologies have been used as long as 25 years ago in the DIANA[5] representation for ADA.

### 2.3 Publish/Subscribe, Actors and Pure Concurrency – No Control of Continuation

There are insights to be gained by examining an extreme example of continuation control – none. This extreme is the model in which tasks are processed in a purely concurrent fashion. By pure concurrency, we exclude the use of constructs like waits, joins, or futures that enable the concurrent tasks to plan for future coordination – the originating task can not even expect to continue to run until the event processing task begins. Unlike stack and workflow models, these models offer the sender or client no reliable control over the concurrent continuation.

We distinguish deliberate lack of control from simple difficulties with delivery QoS, like that encountered in CORBA's "oneway" message delivery specifications[19]. These difficulties ultimately led to the definition of a QoS framework which permits a variety of choices including reliable delivery of one-way messages.[20]. Even assuming reliable delivery, a publish/subscribe model allows a task to call for the performance of other tasks, but not to depend upon their results. Neither the caller nor a global workflow specification controls the continuation. Instead, the continuation is provided by the subscribers that attach to the publication channel.

The use of one-way communication is fundamental to the Actor model for software[11] which eliminates the use of call/return in favor of message sending, entirely bypassing the issue of call-latency. But while actors send their messages directly to other actors, publishers send messages to an open-ended subscriber list, for which there may be many servicers.

The overheads associated with publish/subscribe are quite similar to those associated with workflow models. Neither the originating task nor a workflow manager stores overall state waiting for following tasks, but messages and the objects to which they refer are stored in persistent storage.

While publish/subscribe offers no control to the client, it provides natural flexibility for compositional development. Any event of interest to a concern developer is exposed as a join-point.

### 3. Future Event Handling
#### 3.1 Call-backs
The pure concurrency used in publish/subscribe doesn't provide a naturally useful model for problem decomposition. It is very common for decompositions to result in tasks whose results are required for later tasks. Direct control of the continuation is not available with the pure concurrency of pub/sub. In fact, there is an intrinsic contradiction in the idea of waiting for a task to finish which may not even start until you have finished yourself.

Call-backs are a device by which a client passes to its server some tasks to be performed when the appropriate circumstances arise. But passing the call-back to the service only indicates what "should" happen. If the service fails or has a faulty path it may not call the call-back. Or the coder may simply decide that the circumstances are not appropriate. To use callbacks in problem decomposition, we need a guarantee that the callback will be performed. In the presence of such a guarantee, a problem-decomposer can locally specify a computation in which tasks use the pure concurrency of a publish/subscribe environment yet still exercise local control over the continuation. Providing that guarantee requires us to address three issues: language, supporting environment, and strategies for handling errors.

### 3.2 An Example of Future Commitments

People generally describe problem solutions sequentially, although they can break off chunks described to be done concurrently or in an indeterminate order; and unless prompted, they seldom think of subtasks as subject to long arbitrary delays. So even if a programming language includes an asynchronous message send in addition to a synchronous method call, the feature will be under-exploited because of the difficulty in coordinating the tasks to be performed after the asynchronous task is complete. In consequence, we may expect to see only small improvement in the overall concurrency behavior of software. Unless the concurrent activity can be expressed as pure concurrency ("start it and forget it"), putting it aside breaks the train of sequential thought about the primary problem decomposition and introduces a high intellectual overhead of creating new classes, methods, etc. to deal with the management and coordination of the threaded or otherwise concurrent behavior. To provide both for the better accommodation of surprise occurrences of long latency and for simpler exploitation of concurrency, we need to provide developers with a construct that allows them to think sequentially about activities that can be deferred or executed concurrently. This construct is built on the idea of call-back, with language to provide for a deferred commitment to make the call back, and to allow the passing of deferred commitments from one method to anther. Figure 1 shows two language constructs to provide deferred commitment. They form part of the programming language Continuum/J[22], a Java superset intended to provide more malleable software. While not intending to explore other characteristics of Continuum/J in depth, in reading Figure 1 it should be understood that a Continuum/J `module` is like a Java class, but that the method declarations it contains have no implicit "target" parameter. All parameters are explicit and named. Continuum continues to use the keyword `class` to indicate a syntactic sugaring in which the first parameter remains implicit in method declarations.

In Figure 1, the construct that most directly illustrates use of deferred commitment involves tasks realized as methods called `three` and `four`. It addresses a situation in which the developer knows that after doing some task expressed as an asynchronous method, certain other tasks must be performed. For instance, method `three` might make a bank transfer, for which a receipt must then be sent as proof, via method `four`. This decomposition as "make transfer then send receipt" is a sequential decomposition that would generally be represented by invoking method `three` and then invoking method `four`; or perhaps performing the receipt processing (shown as "…") directly. But in Figure 1, the developer desires to "spin off" the work of task `three`, but does not want to leave the programming language expressing the train of thought or give up all control of the continuation – a situation not addressed by the models described in Section 2.

```
module MyClass {
void one(MyClass this, final Object x, int y)
    sends two(Object x, String z) {
 this.two(x,"any");
 send three(this, 6)
    expect four(MyClass this, int a) {...}
}}
```
Figure 1 – preserving sequence without synchrony

### 3.3 Declaring Future Commitments

To provide local sequential expression in the same language and also control of the continuation, we need to allow the developer to express and ensure a sequential dependency between methods `four` and `three` without holding up the completion of method `one` to do so. Not only does this improve the resource usage in cases of long latency in reaching method `three` but, if we imagine method `one` is called inside a loop then, by allowing it to finish without waiting for method `three` to execute, many executions of method `three` can be started by the loop, all running concurrently. In Figure 1 however, we avoid syntactic interruption of the developer's train of expression by allowing the receipt processing to be guaranteed as part of the call to method `three`, by using a commitment that it will eventually send message `four`.

But the "expect" construct is actually only a syntactic sugar that has three parts:

1. It allows syntactically local encoding of the future work to be performed, embedding it in the body of method.

2. It characterizes the method that performs the work as an *event method*. An event method is a void method that operates in the static or object's scope even though syntactically embedded in another method. Because it is not in the local scope, it has no lexical access to any of the parameters or local variables of that method.

3. It presumes a declaration of `three` expressing a *declarative future commitment* that `three` will cause `four` to execute.

Future commitment is declared as part of an extended method header, like that shown in the header declaring method `one` of Figure 1. In familiar fashion, it declares the method named `one` to be a void method of two parameters: an Object and an integer. However, the "sends" clause extends that declaration to indicate that an invocation of `one` commits to the future invocation of `two`. The clause also indicates an additional constraint. While the names given to parameters is often arbitrary, the use of a parameter in a "sends" clause that has the same name as a parameter in the method declaration indicates that *the same value* must be passed to the committed future method as is passed to the method being declared. We occasionally refer to these as the "passed-through" parameters.

The language rules indicate that such a declaration must be provably true using the simple static propagation of value flow that Java applies to other type-checking situations. In Figure 1, the commitment to call `two` is met by its immediate invocation on block entry. Declaring "x" as "final" ensures that the value is the same as it is on entry.

When checking method conformance, the "sends" clause is treated much like "throws". So, similar type-checking of `three` ensures that the bank transfer method eventually causes message `four` to be sent. Four's implementation is as specified in the "expect" clause, and the receipt is ultimately printed. Recall that method `four` cannot access any of the local state of method `one`, which may have completed long ago. But method `four` can use its parameters to access object state as usual.

The dynamic management of the sent messages to ensure delivery is performed by the execution environment, as described in the next section.

### 3.4 Execution Environment Support

Although bulk of the language is best described as a stack model language, we have in fact introduced elements of a workflow model into the programming language. Declarations of connections between tasks in the workflow are effectively introduced via the "sends" clause. As with workflow models in general, the execution environment must have a component to manage delivery of asynchronous messages. In [9], this is called a "service request broker", which operates as a thread in each VM, working in association with the method call dispatcher. Likewise, the execution environment must provide for active elements that receive and process the sent events. In Continuum languages each these active elements is called a `service`, and the programming language provides for their declaration and for their role in management of a service's relationships to the partial state and behavior of objects it supports[9].

Services each have an independent representation of objects' states. References to objects may be represented as opaque pointers as they are in Java or as conventional data, like Strings, without losing their ability to convey information assuring support for methods. Non-opaque references can be passed between services that are distributed around the network. Method call is cooperative, and each service may add to the behavior resulting from a method call, much like the augmentation of behavior provided by advices in aspect-oriented languages[13]. When all services contributing to a method's composite behavior are present in a single VM, the usual approaches to forming a composite plan that includes them in-line suffice [10],[13]. However when the services are distributed into other VMs or across a network, the response must be handled by an event loop associated with the VM's dispatcher and running in its own thread. Whenever a method call/send is made, the dispatcher performs several actions:

1. If control of the composite plan for this method call/send resides with another VM, pass the invocation to it. If the invocation is a call, the thread is suspended until a response is received.
2. If control of the composite plan for this method call/send resides with this VM, cause the services providing the method's implementations to each be executed in their appropriate VM. Method implementations that belong to this VM may be performed directly or by placing them into the event queue, depending on whether the invocation was by call or send, respectively.
3. A service's method implementation appropriate for a method call is determined at the time the service is compiled, using by Continuum's semantics. These assure that each service's implementation for a method it provides externally is determined unambiguously.

These actions are generally optimized. For example, a simple method call whose single implementation is found in the same VM as its client is dispatched with no overhead above the usual indexed lookup.

Failure to dispatch to an implementation can occur because of network disruption, resource unavailability, or exceptions in required services. In this case, an alternative way must be found to honor the future commitments it implied. This is discussed in the next section.

### 3.5 Handling Exceptions

Provision of any construct, like "send", that decouples future execution but still provides for satisfaction of commitments, must address the problem of exceptions and failures. When a sent message commits to the future invocation of another event, we must define what happens if it fails to do so. Although we statically check to ensure that the future send/call will take place, a logic error may prevent this by causing an exception to be thrown. When an exception is thrown, potentially unsatisfied commitments to send an event are satisfied by sending the committed event in such a way as to trigger the exception immediately on entry to the responding method. An example showing how the exception is processed is shown in Figure 2.

```
module MyClass {
event four(required MyClass this, int y)
  {...}
catch (ContinuityError e)
  { /* code to handle the failure may refer
        to e, this, or y */ }
}
```

Figure 2 – catching a continuity error

This example shows a standalone definition of an event `four` like that expected in Figure 1. The definition of an event provides for a `catch` clause, entered if the event is raised because a method committed to call the event was unable to do so. Within the catch clause, the values of the `catch` parameter and the method's parameters are available. The parameters "passed through", as described in Section 3.3, will have the values passed through from the original call. Parameters not passed through are given the language default, which may be null.

A method used to handle future commitments may, of course, be invoked directly. To ensure that the parameters expected to be passed through to it are always present, those parameters should be declared as "required", indicating that they can not be null references.

## 4. USAGE SITUATIONS

The proposed programming language constructs enable the two changes to programming style proposed in Section 3.2:

1. Software containing points of long latency can be developed and expressed locally, within a single programming language. This can simplify the maintenance required when points of long-latency are introduced after initial design.
2. The greater concurrency that can be obtained by using event sending rather than synchronous invocation for calling subtasks from loops becomes simpler and more manageable than it is with native Java constructs, although there are alternative ways that this advantage can be obtained, as described in Section 6.2.

The constructs do, however, impose restrictions on access to data developed in local variables. So the question arises whether these two advantages can be obtained in practice, rather than simply as a possibility. To help answer this question, we performed a small survey of some existing software known to the authors. The survey will be detailed in Section 5, but to better understand the survey we will discuss here an example illustrating the transformation of an existing piece of software to one using deferred commitment, and the inhibitors to the transformation.

The example is abstracted and simplified from a real system, the PlainWay composition engine[10]. In this program, the command processor's "parse" method contains a processing loop that parses and executes directives. These come in three successive groups: 1) input/out directives, 2) "compose" directives, and then 3) "explain" directives, in that order. The first "compose" directive encountered triggers creation of a compositional universe containing or to contain the specified input and output artifacts. The "compose" directives are sent to this universe to record how the output composite is to be formed. After the last "compose" directive, a method provided by the universe is called to perform the defined composition. Each "compose" directive is independent, and is handled by a call to a method called: executeForm(rationale), internal to the parser. The composition is initiated by a call to the internal method: executeComposition(parseOK,rationale). An abstraction of the code can be found in Figure 3.

A rewritten alternative can be found in Figure 4. In this alternative, the original calls to executeForm and executeComposition are turned into "sends", which run independently. We allow the executeForm methods to run independently, subject to the constraint that when they finish, they must also call reportFormed, to allow the overall work to go forward when ready. The key issue at this point is not the bookkeeping to arrange for the independent processing, but the static declaration that ensures the caller of executeForm will be given control when the formation is complete. This takes the form of a conventional callback,

```
public void parse(BufferedReader inStream, CRRationale rationale) throws IOException {
...
    while /*more characters to process in inStream*/ {
...
        else if ((verb=="compose")) executeForm(rationale);
...
        else if (verb=="explain") {... executeComposition(noErrors, rationale);}
...}
    if ... executeComposition(noErrors, rationale);
    }
}
private void executeForm(CRRationale rationale) {
...
}
public void executeComposition(boolean noErrors, CRRationale rationale) {
...
}
```

**Figure 3 – An abstraction of the "parse" task in PlainWay**

```
public event parse(BufferedReader inStream, CRRationale rationale) throws IOException {
...
    while /*more characters to process in inStream*/ {
...
        else if ((verb=="compose")) {send executeForm(rationale);++count;}
...
        else if (verb=="explain"){... send executeComposition(count,noErrors,rationale);   ...}
...}
    if ... send executeComposition(noErrors, rationale);
    }
}
private event executeForm(CRRationale rationale) sends reportFormed(CRRationale rationale) {
...
send reportFormed(rationale);
}
int arrived=0;int required=-1;
public event reportFormed(CRRationale rationale) {
    send excuteComposition(-1, true, rationale);
}
public event executeComposition(int count, boolean noErrors, CRRationale rationale) {
    if (count==-1) arrived++ else required=count;
    if (required>=0 && arrived==required) ...
}
```

although it never "joins" the original thread. The method called back runs in some other thread, as was described in Section 3.4. The key point being made here, however, is that the deferred commitment guarantees that the "callback" will take place.

The additional code is indicated with a grey background. Four important points should be noted. The first point to is that *parse* is marked as an event method rather than as a void method. The significance of this change is that although the task performed by *parse* will be completed eventually, it may not have been completed by the time control returns to a *parse*'s invoker. Although one can either call or send to a void method, an event method can only be invoked via "send". We transform the void method to an event in recognition of the fact that its calls on reportFormed may exhibit arbitrarily long latencies.

The second point to note is that having made *parse* an event method, we can invoke the task assigned to executeForm via *send*. This means, among other things, that the calls to executeForm found in each iteration of the parsing loop can be executed in parallel and may be executed after the loop completes. However, without further change, execution of the parsing loop may race ahead and call executeComposition before *any* of them have been processed. To prevent that from occurring, we note the third and fourth points at which additional changes are required: 1) executeComposition must also become an event method and, 2) it must get executed after all of the executeForm methods finish. The first requirement is easily satisfied by making it an event method and sending to it instead of calling it. However to satisfy the second requirement, we must guarantee that executeForm reports when its task is finished. This guarantee is supplied by part of the revised declaration of executeForm in which it is committed to call or send a method called reportFormed.

In our example, reportFormed simply sends the message on to a modified version of executeComposition, allowing it to proceed if conditions are appropriate.

The bookkeeping is fairly simple – calls on executeForm are counted and one invocation of executeComposition passes this count when the usual logic indicates it should be called. Other invocations of executeComposition pass a "-1" for that value. When the call with a non-negative value has been made and the count it specifies matches the count of the "-1" values, then its usual code is executed to perform the composition.

The thrust of this transformation is to convert a sequentially described and conceived program into one that is actually described in the workflow diagram of Figure 5

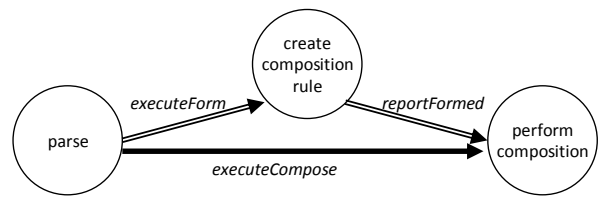

**Figure 5 – Workflow equivalent of Figure 4**

A secondary effect of the transformation is that all the executions of composition rule creations can be run in parallel.

There is also a tertiary effect: that the workflow diagram in Figure 5 becomes the basis for a cooperative workflow structure – one that allows multiplicities of workflows having elements in common to be composed, as symmetric aspects can be composed.

We emphasized that deferring parts of the semantics of a method will generally force us to treat the method as an event. However this need not be so if *join* semantics are needed for compatibility with earlier specifications. Doing so will, of course, remove the advantage of eliminating the costs of maintaining the continuation, but it does promote concurrency of loop behavior. By committing the final activity in the workflow to call an event on its completion, the invoking task can provide an implementation for that event which uses a conventional mechanism like wait/notify to wait for completion. For example, if the declaration of executeComposition in Figure 4 had specified

"`sends compositionComplete(CrRationale rationale)`"

then the parse method could wait for an event and be notified of that event by the implementation of the compositionComplete method's implementation.

## 5. USABILITY SURVEY

To determine whether the restrictions on use of local variables greatly interfere with the applicability of deferred commitment, we examined a small sample of code from an existing body of software known to the authors – the JikesBt concern assembler and the concern composition component from the CME[2]. The sample consisted of 61 classes, in which there were 569 methods.

The survey examined only "void" methods since we conservatively took the return of a value to indicate that a method needed to complete before returning. The assumption is conservative because even value-returning methods may spawn tasks that could complete later. There were 117 non-empty void methods in the sample.

A more detailed discussion of the survey will be made in the next two sections. In summary, of the 117 void methods, 69 were deemed to be reasonably made into event methods if their callers were suitably modified. Of those 69, 34 were manifestly so and 35 others were judged so by two reviewers on the basis of a semantic understanding of the method. The 117 void methods contained 43 loops of interest, of which 38 were judged to be expressible as concurrent event methods. Of these 38, in 31 cases the conclusion was manifest, and the 7 others required semantic understanding of the method to reach the conclusion.

### 5.1 Analysis of Event Methods

Transformation of a void method to an event method is not allowed if the overall result is changed significantly when invocations of the method are executed in an order different from that in which they are called. We do not regard aspect semantics like logging to be significant in this respect. Interaction between invocations of a method can occur when the method both uses and changes the value of the object's fields, either directly or through another method call. In methods that make no calls, such interaction is manifestly absent if the method refers only to fields or static variables, or to no such variables at all.. This can easily occur if the primary reason for making the code into a method is because polymorphism is being used to select an algorithm, and not to access object state.

Where the method makes calls, a semantic understanding is applied to judge whether those calls induce an order dependency that would inhibit transformation. To strengthen the results, judgments were made by two analysts, and the method was considered a candidate only if both agreed.

Applying these criteria, we examined the 117 non-empty, void methods in our sample to see if they could have been written as event methods. Of these, 34 met the criteria identified above for manifest potential. In addition 35 others were judged to have the potential, by two reviewers on the basis of a semantic understanding of the method. The remaining 48 were not deemed acceptable.

Each point at which a method is called adds to the management overhead involved in treating it as an event, unless these call points cluster near the end of the logic containing the call. We listed the methods of interest that invocation point counts suggested might incur too much effort, but none of the methods we had identified as candidates for converting to events fell into this list.

### 5.2 Analysis of Loops with Concurrency

In addition to examining whether void methods might be turned into events with deferred commitments, we examined the same suite of classes to determine whether the material in the loop bodies could be run concurrently using deferred commitments. The criteria for determining whether this transformation is possible are similar to those used for examining event methods in the prior section: avoiding situations where the portion to be parallelized both uses and changes the value of the object's fields, either directly or through another method call. In addition, method extraction, used for transforming part of a loop interior into a method call, has a number of preconditions that must be satisfied, including one that the method body may not assign to local variables outside its scope.

We discounted as "uninteresting" those loops in which each iteration did not appear to contain enough "work" to overcome the costs of transforming the content into a separately called method. There were 43 interesting loops, of which 38 were judged to be expressible as concurrent events using a counting mechanism like that illustrated in Figure 4 to synchronize for continuation. Of these 38, in 31 cases the conclusion was manifest. The 7 others required semantic understanding and were judged transformable by two reviewers on the basis of a semantic understanding of the method.

## 6. RELATED WORK

We have proposed a programming language construct useful for application to two common problems: 1) reducing the run-time overheads for retaining continuations at points of long latency and 2) simplifying the introduction of increased concurrency in loops that can be parallelized.

### 6.1 Long Latency

Rather than looking at support for long-latency calls from a second-language or extra-lingual point-of-view, as is done with workflow paradigms or with framework support like Enterprise Java Beans or its J2EE successor, we have focused on internalizing its support in programming language features. Although concern for long-latency operations is present in web applications, the paradigm-breakage introduced in workflow and the run-time overheads of its conversions to/from XML limit its applicability and require an investment in specialized training. The primary interest in language support for long-latency operations comes from sensor-network and embedded application environments.

In this context, TaskJava[4] characterizes the domain as one of "event-driven" programming, introduced with the characterization: "Event-driven programming implements a stylized programming idiom where programs use non-blocking I/O operations, and the programmer breaks the computation into fine-grained callbacks (or event handlers)". But, in one sense, this description presupposes an answer to the question of "who controls the continuation" posed in Section 2, asserting that control belongs to the writer of a method instead of the writer of the task assigned to the method (using the distinction made in Section 1). This writer is burdened with the fact that callbacks represent an inversion of control. If, as we suggest here, the discussion of event handlers is framed in terms of the pub/sub model, the continuation does not belong to any single author, and there is no inversion of control. There is simply a need for the writer of one task to assure that a subtask will be performed in the future.

The basis concept for the language features suggested here is that the use of a broadcast model of method call enables us to introduce declarative structures that provide the method-writer with a way to participate in the overall definition of the work-flow of events. This enables the writer to *share* control of the continuation without being *in control of* it. Instead of treating "inversion of control" as an awkward problem, we treat "loose control" as an advantageous norm.

This is quite unlike the approach taken in TaskJava, which is also concerned with supporting event programming. TaskJava introduces the language construct "Task" (which is capitalized here to avoid confusion with the use of the term *task* in this paper generally). In TaskJava, Tasks replace threads as the unit of concurrency, but communication is still target-directed. On the other hand, in a broadcast-model, task programmers are concerned not with organizing the units of concurrency but with controlling the sequence of events. Concurrency management lies with the Services, described in Section 3.4. So, while new tasks are spawned explicitly by a the client in TaskJava, they are spawned as a hidden part of service management with the broadcast model of call.

With declared future commitments, the task writer is concerned with isolating elements that are potentially separated by long-latency operations, whereas in TaskJava, the points of potential latency are inferred by compiler. But this automatic inference encourages careless carriage of state across the points of latency. The use of declared future commitments enforces a natural linguistic barrier to automatic carriage of state. However, this then requires deliberate thought about where a task should be divided into different events. The formulation of TaskJava notes that the continuation of a method need not be put in a separate class, and observes that this locality presents advantages over common call-back patterns that require an extra class to hold the call-back. Declared future commitments provide an alternative middle-path: in which the continuation is a new method within the original class but syntactically nested in the original method.

These differences suggest not a superiority of one approach over the other, but rather that different issues and values are being addressed by them.

The event-loop structure of services in the broadcast model for calls evokes similarities to the structure of actors, used to provide event-based software without inversion of control in Scala[6]. The description of this approach concludes that "… we kept to a large extent the programming model of thread-based actors, which would not have been possible if we had switched to a traditional event-based architecture, because the latter causes an inversion of control." The implicit claim is that the actor model naturally avoids inversion of control. But they discuss no provision for allowing the dissection of a task across several actors with a static typing mechanism to enforce the satisfaction of locally known needs for particular behavior in the continuation.

The FlowTalk programming language[1] also addresses the issue of long-latency operations. Like TaskJava, the compiler cuts methods into fragments that can be scheduled separately. FlowTalk avoids issues concerned with holding storage resources over long-latency calls, addressing storage resource limitations by, for example, avoiding dynamic object creation. The use of automatic division avoids the need for declaration and enforcement of future commitments, but suffers from the difficulties of hidden overheads and lack of composability alluded to earlier. But explicit splitting of the tasks requires an enforcement mechanism for future commitments.

## 6.2 Concurrency

Call-backs are a well-known mechanism by which threads created to carry out subtasks of some target task can report their results or completion. In object-oriented software, they are often passed to the subtask's thread as methods in a parameter of the thread's creation. Figure 4's use of `reportFormed` is an example of a callback to be used by `executeForm`. It is passed as one of the public methods in the *this* object passed to `executeForm`. However, in the usual formulation of callbacks, there is no way to ensure that the callback to report results of completion will always be made. Declarative future commitment provides a language construct for declaring and enforcing that the callback will be made by the implementation of the sub-task.

As an alternative to using callbacks to manage concurrency, programming languages like E[17] provide a language construct called Futures. Instead of calling back, the sub-task assigns a value to the future variable. If and when the primary task wishes to use the result, it retrieves the value. All waiting is hidden by the language support. Futures provide a simpler, more directly expressive mechanism when the super-task wishes to wait for one or more results before proceeding, the case illustrated in Figure 4. But if the super-task wishes its continuations of subtasks' work to be themselves concurrent, callbacks provide the more versatile primitive. In this circumstance, declarative future commitments make it possible to guarantee that callbacks will be called.

## 7. SUMMARY

We began this paper with an exposition of several different models for expressing the decomposition of tasks into subtasks and for controlling the future flow of control among them, noting that they provide very different answers to the question of who controls the continuation. This exposition focused attention on the tension between local expression of the sequence of tasks in a single program and the inflexibility and costs of such local expression when faced with long-latency communication between sub-tasks. We described a language construct, called declarative future commitment which can be used to ensure that tasks can be expressed locally but guaranteed to be carried out eventually in the context of a broadcast model for method call. We then reported that a small study indicated that this feature could be usefully exploited in a significant number of instances both for managing possible long-latency operations and for improving the reliability and expression of concurrency in subtasks occurring in loops.